# Atomic Structures of Graphene, Benzene and Methane with Bond Lengths as Sums of the Single, Double and Resonance Bond Radii of Carbon

Raji Heyrovska

Institute of Biophysics, Academy of Sciences of the Czech Republic.

Email: rheyrovs@hotmail.com

## Abstract

Two dimensional layers of graphene are currently drawing a great attention in fundamental and applied nanoscience. Graphene consists of interconnected hexagons of carbon atoms as in graphite. This article presents for the first time the structures of graphene at the atomic level and shows how it differs from that of benzene, due to the difference in the double bond and resonance bond based radii of carbon. The carbon atom of an aliphatic compound such as methane has a longer covalent single bond radius as in diamond. All the atomic structures presented here have been drawn to scale.

## Introduction

A good introduction to graphene with diagrams and literature survey of the field can be found in [1,2]. It can be seen from [1] that research on graphene is currently pursued in many scientific institutions, and notably by the Geim group [2]. The structure of graphene is presently represented as a ball and stick model [3]. Since both benzene and graphene involve hexagonal arrangement of carbon atoms, a deeper insight into their atomic structures is provided here. Based on the author's previous work on the atomic structures of moleclues with bond lengths as sums of atomic covalent radii [4-7], it is shown here that the bond lengths in the above molecules can also be represented as sums of the appropriate atomic radii of the carbon atoms involved.

The atomic covalent radius of carbon, defined as half the covalent bond lengths [8], in diamond and aliphatic compounds is known to be different from that in graphite and aromatic rings [8], which involve resonance bonds. The author has shown [4-7] that the atomic radius of carbon in graphite holds for three alternate atoms in the benzene ring, while the other three carbons have the covalent double bond radius, thus accounting for the C-C bond length in benzene.

## Atomic radii and bonding in methane, benzene and graphene

It is shown here in Fig. 1 (top left) that diamond and aliphatic compounds [8] like methane involve $C_{s.b.}$ with the single bond (subscript s.b.) covalent atomic radius of 0.77Å. In methane, the $C_{s.b.}$- H distance is the sum, $0.77 + 0.37 = 1.14$Å, where the covalent radius of H is 0.37Å. This is close to the C-H distance in [8] of about 1.10 Å.

Benzene (Fig. 1, top center) has six C-C bonds of equal distance [8] of 1.39 Å (+/-0.02), which was shown [4] to be the sum of the atomic covalent double bond (subscript: d.b.) radius of $C_{d.b.}$ (0.67 Å) and of atomic resonance (subscript: res.) bond radius of $C_{res.}$ (0.71 Å), [4-7]. Note that $0.71 \sim (0.77 + 0.67)/2 = 0.72$ Å, as if the resonance bond has a 1.5 bond character [8]. The six ring carbon atoms are bonded covalently to six H atoms through six single bonds. Note that there are two C-H bond distances in benzene (not recognized before): $C_{d.b.}$- H = $0.67 + 0.37 = 1.04$ Å and $C_{res.}$ - H = $0.71 + 0.37 = 1.08$ Å. The value in [8] is 1.08 Å. The center of the benzene ring can fit exactly another $C_{d.b.}$ of radius 0.67 Å, as shown by the empty inscribed circle.

The graphene hexagon (Fig. 1, right) has a C-C bond length [8] of 1.42 Å which is twice the resonance bond atomic radius of $C_{res.}$ (0.71 Å), and occupies an area of about 4.3 x 4 Å, which is 0.43 x 0.4 = 0.17 $nm^2$. The six ring carbon atoms have six free bonds (two resonance bonds and four single bonds, see Fig. 1, right) which covalently bind to six other carbon atoms (not six hydrogens with single bonds as in benzene), forming six other rings of graphene as shown in Fig. 2. The center of any hexagon in graphene can fit an inscribed circle (see the empty circle in Fig.

1, right) equal to that of another $C_{res.}$ of radius, 0.71 Å, whereas benzene can fit a smaller circle with $C_{d.b.}$ of radius 0.67 Å (see Fig. 1, center).

Thus, the author has presented here the atomic structure of graphene and has shown how the radii and bonding of carbon differ from that of benzene, although they both involve hexagons. Methane and other aliphatic molecules involve $C_{s.b.}$ as in diamond.

**Acknowledgements:** The author thanks the IBP for financial support and Professor A. Geim, F.R.S. of the University of Manchester, for sending me the reference [2] containing many downloadable articles.

**Fig. 1** Atomic structures of methane, benzene and graphene (0.43 x 0.4 nm).

Methane                    Benzene                    A graphene hexagon

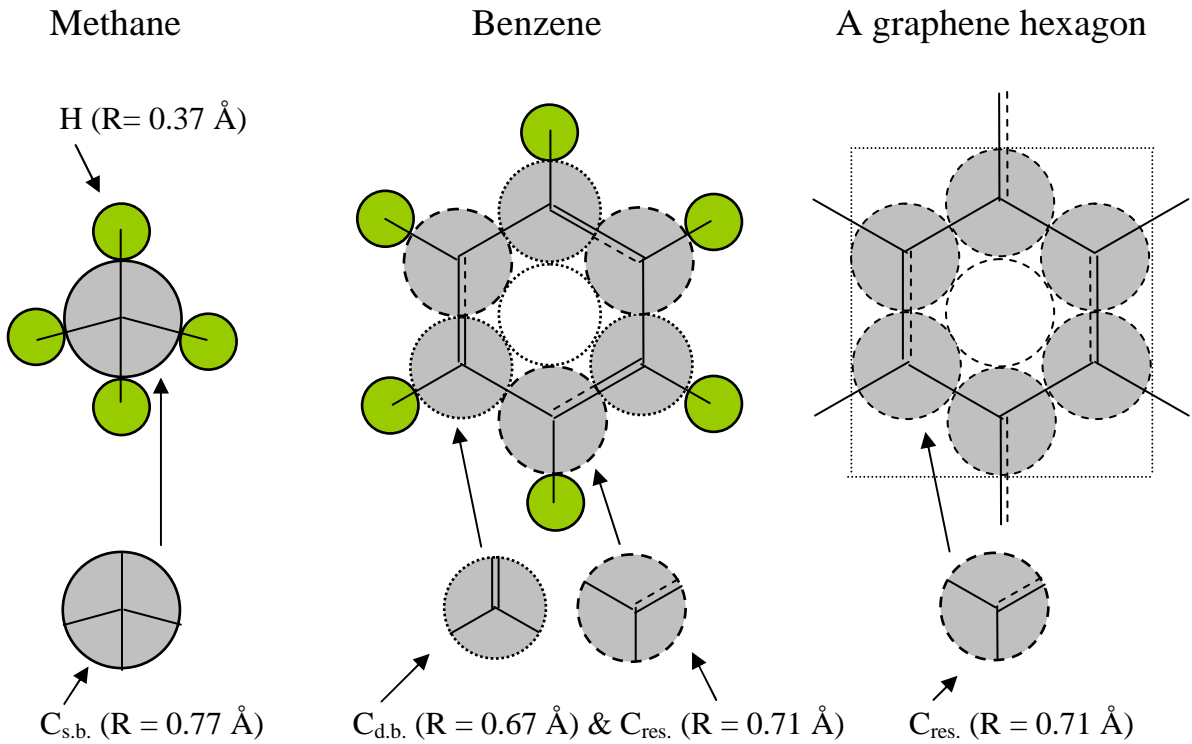

H (R= 0.37 Å)

$C_{s.b.}$ (R = 0.77 Å)

$C_{d.b.}$ (R = 0.67 Å) & $C_{res.}$ (R = 0.71 Å)

$C_{res.}$ (R = 0.71 Å)

**Fig. 2** Atomic structure of graphene.

24 carbon atoms making 7 hexagons in 0.9 x 0.9 nm (~ 0.8 nm$^2$) of graphene.

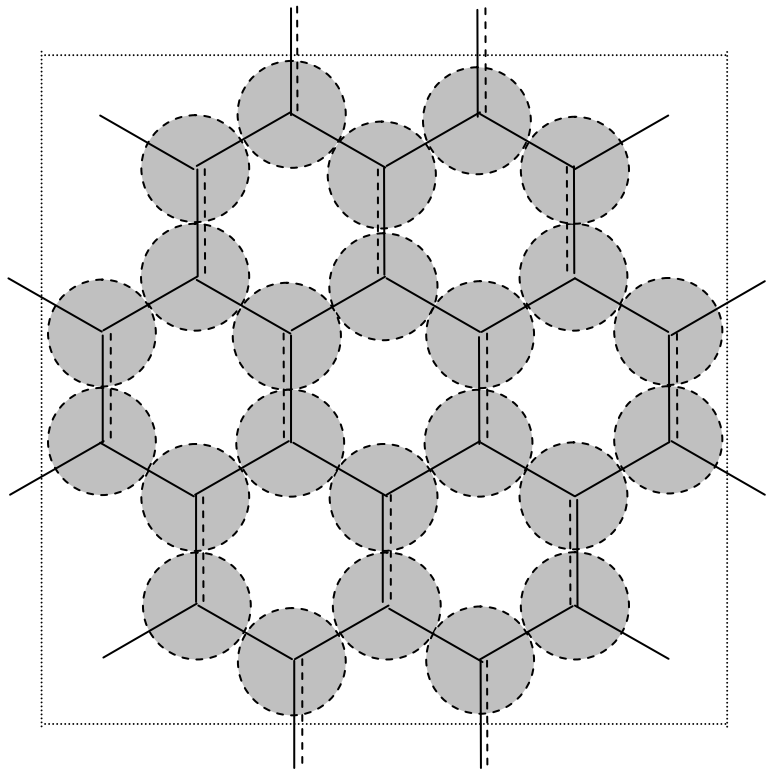